\documentclass[12pt]{iopart}
\usepackage{graphicx}
\usepackage{subfigure}

\begin{document}

\title[]{The precursor of the critical transitions in majority
vote model with the noise feedback from the vote layer}

\author{Wei Liu$^{1,2}$, Jincheng Wang$^1$, Fangfang Wang$^{2,1}$, Kai Qi$^3$ and Zengru Di $^2$ }

\address{$^1$College of Science, Xi'an University of Science and Technology, Xi'an, China}
\address{$^2$International Academic Center of Complex Systems, Beijing Normal University, Zhu Hai, China}
\address{$^3$2020 X-Lab, Shanghai Institute of Microsystem and information Technology,
Chinese Academy of Sciences, Shang Hai, China}
\ead{weiliu@xust.edu.cn \\
     kqi@mail.sim.ac.cn\\
     zrdi@bnu.edu.cn}
\vspace{10pt}
\begin{indented}
\item[] 
\end{indented}

\begin{abstract}
In this paper, we investigate phase transitions in the majority-vote model coupled with noise layers of different structures. We examine the square lattice and random-regular networks, as well as their combinations, for both vote layers and noise layers. Our findings reveal the presence of independent third-order transitions in all cases and dependent third-order transitions when critical transitions occur. This suggests that dependent third-order transitions may serve as precursors to critical transitions in non-equilibrium systems. Furthermore, we observe that when the structure of vote layers is decentralized, the coupling between the vote layer and the noise layer leads to the absence of critical phenomena.

\end{abstract}

\vspace{2pc}
\noindent{\it Keywords}: Majority-vote model, Multi-layer networks, Critical transitions,
Third-order transitions
%\maketitle

\section{\label{sec:level1}Introduction}

Phase transitions exist widely in nature. From
traditional materials to the biological flock, both of equilibrium and
non-equilibrium transitions, particularly critical transitions, exhibit similar behaviors within the same universality classes \cite{pathria2011statistical,vicsek2012collective}.
The transition from a disordered configuration
to an ordered state in social systems such as opinion formation, cultural dynamics,
language dynamics has attracted much attention in recent years as well \cite{castellano2009statistical,dorogovtsev2008critical}.
Near the critical point, lots of systems including social systems exhibit a quick response to
external disturbances and avalanches. When the control parameters exceed their critical values, large-scale social disturbances occur \cite{munoz2018colloquium}. Hence, predetermination of the critical
points or the locations of the points should be of significance.
Statistical physics has proven to be a very powerful
framework to describe the phenomena and can help us well understand the
essence of the critical transitions.

Recently, Qi and Bachmann \cite{qi2018classification} generalized the microcanonical
inflection point analysis method \cite{gross2001microcanonical} to identify and locate independent and dependent
phase transitions. The microcanonical entropy and its derivatives are
monotonic functions
within energy regions associated with a single phase. However, a phase
transition would break the monotonicity and is singled out by an inflection point.
Independent transitions resemble traditional transitions and occur independently of other collaborative activities within the system. In contrast, dependent transitions are always connected to an independent transition. These transitions happen at higher energy levels and are of a higher order than the independent transitions they accompany.
This method has been applied to the study of Ising model,
flexible polymers and Baxter-Wu models, and has proven to be successful in 
these models \cite{qi2019influence,liu2022pseudo,sitarachu2020exact}.
The results of these studies reveal major transitions and distinguish the details of the transition
processes by signaling higher-order transitions. Moreover, they found dependent transitions that
can only occur in co-existence with independent transitions of a lower order.

It has proven useful as a foundation for a better understanding of general geometric properties of phase
transitions \cite{pettini2019origin,bel2020geometrical} as well.
K. Sitarachu and M. Bachmann \cite{sitarachu2022evidence} studied the Ising model in square lattice
and found that the average cluster size (ACS) (excluding isolated single spins)
becomes extremal at about the temperature
of the third-order dependent transition in the paramagnetic phase. 
The decrease of ACS with increasing
temperature is expected in the paramagnetic phase. This decrease accelerates for temperatures
lower than the third-order dependent transitions, before slowing down for temperatures
higher than the dependent transitions. This unexpected system
behavior challenges the assumption that ACS would
decrease monotonically in the disorder phase, potentially eliminating
the third-order dependent transition. Despite appearing as a minor effect, 
this shift in monotonicity serves as a crucial indicator of the catastrophic
critical transition. 
In the ferromagnetic phase, the average number of isolated spins (ISN) at the independent third-order transition
temperature, which was identified by microcanonical analysis, has local minima. Here, the increased number of such 'seeds' of disorder in the
ferromagnetic phase enables the formation of critical clusters once the critical point is approached.

Can the two new order parameters (ACS and ISN) be applied to non-equilibrium complex systems
to forecast the critical transitions? After S. Galam's pioneer work
\cite{galam1986majority,galam1991towards,galam1997rational} on opinion formation, lots of studies \cite{krapivsky2003dynamics,sznajd2005sznajd,sznajd2004dynamical,sastre2016antiferromagnetic}
have emerged in this field. One well-known non-equilibrium model in this domain is the isotropic majority vote (MV) model. This model exhibits a continuous order-disorder phase transition and has demonstrated that non-equilibrium stochastic systems with up-down symmetry belong to the universality class of the equilibrium Ising model \cite{de1992isotropic}. The vote models in complex
networks \cite{lima2012nonequilibrium,alves2021consensus,stone2015majority,chen2015critical,krawiecki2018spin,choi2019majority}
have been investigated and many of them focus on the effects of underlying topologies,
such as random graphs \cite{pereira2005majority,lima2008majority}, scale-free networks \cite{lima2006majority}, small world networks \cite{campos2003small,luz2007majority} and other complex
networks \cite{santos2011majority,fronczak2017exact,lima2006majority,wu2010majority,lima2013majority}.
The critical transition is observed in different
types of network but with varied critical noise $q_c$ and critical exponents \cite{chen2015critical}.

However, all the aforementioned works overlook a crucial aspect present in real social systems—namely, the interaction between systems and their environment, where mutual influence is exerted. The environment, similar to social noise, can significantly affect decision-making \cite{pampapura2022impact}. This noise has the potential to diffuse into a stationary and homogeneous state. Various forms of social noise, such as rumors, misinformation, or unpredictable events, may propagate their influence throughout a community or society. Recent studies have delved into the diffusion of social noise \cite{del2016spreading, yang2020multiplexity}, with authors emphasizing homogeneity as a key driver for the diffusion process \cite{del2016spreading}. Modeling the interaction between social systems and their environments through multilayer networks emerges as a natural and apt choice \cite{perc2019diffusion}.
Of particular interest in the study of multilayer networks is the coevolution between different dynamical processes. 
Liu et al studied a non-equilibrium model known as the majority vote model
coupled with reaction-diffusion processes on a two-layer multiplex network \cite{liu2019coevolution}.
Their coupling mechanism induces a continuous order–disorder phase transition on random
regular graphs, although this critical phenomenon disappears on square lattices.
It needs to be ascertained whether the occurrence of this critical transition depends on the network topology or the feedback mechanism.

We will compare the behaviors of the MV model with different type of noises:
homogeneous one caused by external environment, another homogeneous one caused
by the feedback from the vote dynamics and heterogeneous noise coupled with the
vote layer. It should be interesting to find signals to forecast the critical transitions in
the systems and to reveal which mechanism (topology or feedback) is more
important for critical transitions.
In Section \ref{sec:model}, we provide a brief description of the model and our methodology. The results are presented in Section \ref{sec:results}, followed by discussions. Section \ref{sec:summary} concludes the paper and outlines potential future directions.

\section{\label{sec:model}The model and the method}
\subsection{The model}
\begin{figure}[htbp]
\centering
\includegraphics[width=0.55\textwidth]{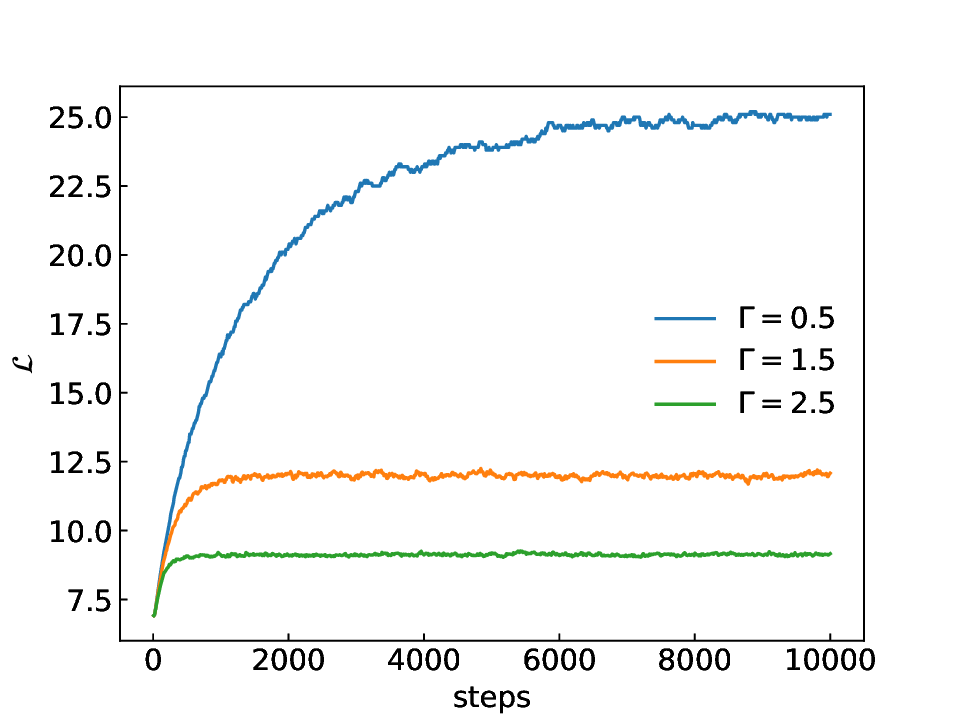}
\caption{\label{fig:aspl} The average shortest path length.
}
\end{figure}

\begin{figure}[htbp]
\includegraphics[width=0.9\textwidth]{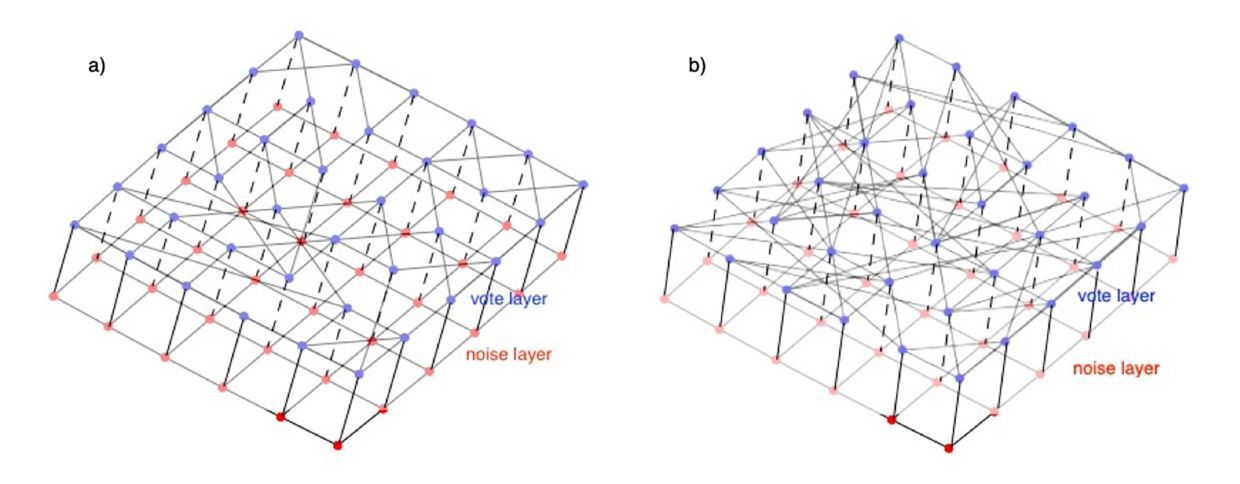}
\caption{\label{fig:2-l} The bilayer complex network. The upper layers are the
vote layers, which are RRNs with a) $\Gamma = 0.5$ and b) $\Gamma = 2.5$, and the lower layers are the noise layers.
}
\end{figure}

Our model is defined on a two-layer multiplex network, with each layer initially arranged in a $N \times N$ square array. The structure of the network's links is upon the specific network types employed.
Square lattices (SL) and random regular networks (RRNs) are utilized in this work. To study
the effect of the decentralization of the network on the phase transitions, we rewire the
links of the RRN to enlarge the average shortest length \cite{schrauth2018two} of the specific layer. The parameter $\Gamma$  (the
characteristic inverse shortest distance) is introduced to govern this decentralization effect. For rewiring, pairs of connected edges, denoted as $(i, j)$ and $(k, l)$, are chosen at random. When the two edges are cross-rewired, giving rise to new connected edges,
$(i,l)$ and $(j,k)$. The Euclidean distances, denoted as $d(i,j)$, $d(k,l)$, $d(i,l)$
and $d(j,k)$ prior to the rewiring are calculated. Then the cost function $\Delta d$ is defined,
\begin{equation}
\Delta d \equiv d(i,j) + d(k,l) - [d(i,l) + d(j,k)].
\end{equation}
If $\Delta d < 0 $, the change is always accepted. Otherwise, the rewiring is accepted
only with probability 
\begin{equation}
P = \exp( - \Delta d/\Gamma).
\end{equation}
It is crucial to emphasize that the parameter $\Gamma$ has a substantial impact on the average shortest length of the RRNs. Fig. \ref{fig:aspl} illustrates the average shortest path length $\mathcal{L}$ as a function of the rewiring steps of the links. After a sufficient number of rewiring steps, the average shortest path length stabilizes.
Refer to Fig. \ref{fig:2-l} for a graphical depiction of our model, which elucidates that higher values of $\Gamma$ result in shorter average shortest paths. Through the manipulation of $\Gamma$, we can craft network structures with adjustable average shortest path lengths. Although each node in a
RRN is equivalent, the shorter average shortest path
implies that each node is more likely to be the center of the system. In
contrast, in a 2D SL, the average path length is longer and each point
cannot be the center of the system. This results in a reduced effective
dimension of the random regular network compared to the 2D regular lattice,
as we conclude later in the section of the results on the behavior of the MV
model. The change in the effective dimension would give rise to variance in critical behaviors.

Consider a complex network in the upper layer (vote layer), where each node is associated with a spin variable denoted by $\sigma_{i}^{V} = \pm 1$ and
the configuration of the system denoted by $\{\sigma_{i}^{V}\}$.
A given spin flips with
probability $q_{i}^{N}$ if it agrees with the majority sign
and flips with probability $1-q_{i}^{N}$ if it does not.
The origin of the parameter $q_{i}^{N}$ could be attributed to various sources: it might emanate from the external environment, emerge as a feedback effect from the entire voting system, or result from feedback generated within the local cluster accompanied by diffusion on the noise layer. The subsequent paragraph will delve into elaborations on the latter two scenarios.
The spin-flip probability is then defined by
\begin{equation}\label{flip_pro}
  w(\sigma_{i}^{V}) = \frac{1}{2}[1-(1-2q_{i}^{N})\sigma_{i}^{V}S(\sum_{j}a_{i,j}^{V}\sigma_{j}^{V})],
\end{equation}
where $S(x) = \rm sgn (x)$ if $x \neq 0$ and $S(0) = 0$. The summation is over the neighborhood of the site and
$a_{ij}^{V} = 1$ if site $i$ and $j$ are linked in the vote layer. The
superscript $N$ in the formulas represents the variable of the noise layer,
while the superscript $V$ represents the variable of the voting layer.

The noise parameter should reflect
the degree of a person’s rationality. It describes social tension that affects human
behaviors. Liu et al \cite{liu2019coevolution} utilized a reaction-diffusion equation to describe the evolving
of the noise in another layer of the multiplex networks.
\begin{equation} \label{noise_ev}
    \frac{dq_{i}^{N}}{dt} = -a + b\frac{C_{i}^{V}}{N} + D\sum_{j=1}^{N}L_{ij}^{N}q_{j}^{N}.
\end{equation}
The parameter $a$ captures a reduction in noise. The second term signifies the alleviation of societal tension due to the formation of collective structures within the voting layer. As clusters form, governmental interventions are prompted to mitigate societal tension and prevent further civil disturbances. Here, $C_{i}^{V}$ corresponds to the cluster size of site i in the voting layer, and $N$ represents the total number of nodes.
The last term accounts for the diffusion of the noise within the noise layer. $D$ stands for
the diffusion coefficient, and $L_{ij}^{N}$ is the Laplacian matrix,
defined as $a_{ij}^{N}-k_{i}^{N}\delta_{ij}$ where $k_{i}^{N}$ is the degree of the node $i$ in the noise layer.

If we consider the feedback without diffusion, all $q_i^{N}$ have the
same value $q$. The Eq. (\ref{noise_ev}) can be
derived to
\begin{equation}\label{noise_wod}
   \frac{dq}{dt} = -a + b[|m| + C_{max}^{V}/N],
\end{equation}
where $m$ is the average of the spin and $C_{max}^{V}$ is the maximum cluster size.
In the thermodynamic limit, the Eq. (\ref{noise_wod}) enters the mean-field regime.

The competing of the two mechanisms represented by the ratio $r = a/b$ (competing ratio), governs the trajectory of the system's evolution, with a predetermined value of $a$ held constant. If noise parameters in the noise layer become too small,
the formation of clusters in vote layer will tend to increase it. On the other hand, if noise parameters
become too large, the cluster size will also be small and the constant decay will tend to decrease it.
As a result of this delicate balance, a combination of constant decay and the influx of external noise passing through the noise layer clearly steers the system toward a non-equilibrium steady state.

\subsection{The method}

We carry out extensive Monte Carlo (MC) simulations on RRN
and SL. The vote layer and the noise layer are not required to be the same. We investigate
the different combinations: Both layers are RRNs or SL, or one is RRN and the
other is SL.
For the coupled evolving situation, we first update the configuration of vote layer
and then update noise parameters in noise layer. Considering different time scales of the
different dynamical processes, we update the spin configurations according to Eq. \ref{flip_pro},
then update noise parameters according to Eq. \ref{noise_ev} or \ref{noise_wod} after
$\tau$ MC steps. Typically $5 \times 10^5$ to $5 \times 10^6$ MC
steps per site (MCS) were discarded and the same MCS were
retained for the averages. $\tau$ equals $5$ to $20$ depended on
the structures of the noise layers in our simulations. The system sizes $N$ are taken from $16$ to $64$ for multilayer networks of which both layers are RRN or for single layer networks.
However, $N = 16$, $24$ and $32$ if one layer of the bilayer networks is SL. The
reason for taking these system scales is that when a layer is SL, 
the dynamical processes of that layer take longer to reach stability compared to
RRN. This results in very long computational time
in simulations. The noise diffusion reaches a steady state after $D_t$ steps. Therefore, the the total simulation time should be proportional to $D_{t} N^2 $.
In our simulation, $D_{t} = 50$ when the noise layer is RRN and $D_{t} = 500$
when noise layer is SL.
The measured quantities in our simulations are the magnetization
$M$, mean value of noise
$\overline{q}$, susceptibility $\chi$ and Binder cumulant $U$ at different ratio $r$:
\begin{equation}
    M = \langle|m|\rangle = \langle\frac{1}{N^2}|\sum_{i=1}^{N}\sigma^{V}_{i}|\rangle,
\end{equation}
\begin{equation}
    \overline{q} = \langle \frac{1}{N^2}\sum_{i=1}^{N}q^{N}_{i}\rangle,
\end{equation}
\begin{equation}
    \chi = N^2(\langle m^{2}\rangle- \langle m \rangle^{2}),
\end{equation}
\begin{equation}
    U_4 = 1- \frac{\langle m^4 \rangle}{3\langle m^2\rangle^2}.
\end{equation}
For the noise coming from external environment, standard MC simulations are
carried out and $\overline{q}$ is not calculated. The other quantities are
obtained at different noise $q$.

To obtain the critical exponents, we utilize finite-size scaling theory \cite{landau2021guide}.
According to this theory, the thermodynamic properties
obey the scaling forms, e.g.,

\begin{equation}
  M\propto N^{-\beta/\nu},\label{magnetic}
\end{equation}
\begin{equation}
  \chi \propto N^{\gamma/\nu},\label{susceptibility}
\end{equation}
where $\beta$, $\gamma$ and $\nu$
are critical exponents that should obey the scaling relation at second-order phase transitions.
The exponents $\beta$, $\nu$ and $\gamma$ should satisfy the scaling relation:
\begin{equation}
    2\beta + \gamma = \nu D_{eff}. \label{realtion}
\end{equation}

To determine the critical transition point accurately, the location of peaks in bulk quantities
defines an effective transition ratio or noise that varies with the system size as

\begin{eqnarray}
    r_c (N)& = &r_{c} +\lambda N^{-1/\nu} \nonumber \\
    q_c (N) & = &q_c +\lambda N^{-1/\nu},  \label{criticalt}
\end{eqnarray}
where $r_{c}$ is the inverse critical ratio at the thermodynamic limit and $\lambda$ is a constant.
Obviously, the value of $\nu$ is necessary to obtain the other critical exponents and the critical temperature at $N\sim \infty$.
We use the relation of the maximum of the derivatives of $U$ to obtain the exponent $\nu$:
\begin{eqnarray}
       \frac{\partial(U_4)}{\partial q}|_{max} &\propto &N^{1/\nu} \nonumber \\
       \frac{\partial(U_4)}{\partial r}|_{max}&\propto& N^{1/\nu}. 
\end{eqnarray}

To detect the higher-order transitions, two extra parameters should be calculated during
the simulations. They are ACS and ISN which give the signals of the dependent and independent
third-order transitions, respectively. Sitarachu et al define $A$ as the ACS containing more
than a single spin in a given spin configuration $\{ \sigma \}$
\begin{equation}
    \langle A \rangle = \langle \frac{1}{n^{\prime}}\sum_{ l^{\prime}}C_{l^{\prime}}\rangle,
\end{equation}
where $l^{\prime}$ labels the clusters with more than one spin, $C_{l^{\prime}}$ is the number of spins in
cluster $l^{\prime}$, and $n^{\prime}$ is the total number of clusters with more than one spin in
$\{ \sigma \}$. $\langle ...\rangle$ denotes the statistical average taken over $10^5$ - $10^6$ MCS.

$d \langle A \rangle/dT$ exhibits a local minimum at the dependent third-order transition point for
the Ising model. This transition takes place in the paramagnetic phase, in which
the average cluster size decreases with increasing temperature. This decrease accelerates for
temperatures below the dependent third-order transition point while slows down above
the point. The unexpected minor change of monotonicity is an important signature of
the catastrophic critical transition.
The third-order independent transition occurs in the ferromagnetic phase.
It reflects the emerging disorder and entropic variability. We use the
single-spin cluster, i.e., an isolated single spin surrounded by nearest-neighbor
spins with opposite orientation which suggested by Sitarachu et al.
The statistical average of the number of isolated spins per site $\langle n_{1}\rangle$ 
shows a local maximum at the independent third-order transition point.

The positions of the two types of third-order transitions determined by $d\langle A\rangle/dT$ and $\langle n_{1}\rangle$ align with the outcomes from the microcanonical inflection point analysis in the Ising model. Both the MV model and the Ising model belong to the same universality class \cite{castellano2009statistical,de1992isotropic}. The divergence behaviors of the correlation length in these models are identical, suggesting a similar cluster formation process near the critical point. The parameters signaling the third-order transitions in the Ising model could be effectively applied to the MV model.

\section{\label{sec:results}Results}
\subsection{Traditional critical transitions}

\begin{figure}[htbp]
\centering
\includegraphics[width=0.85\textwidth]{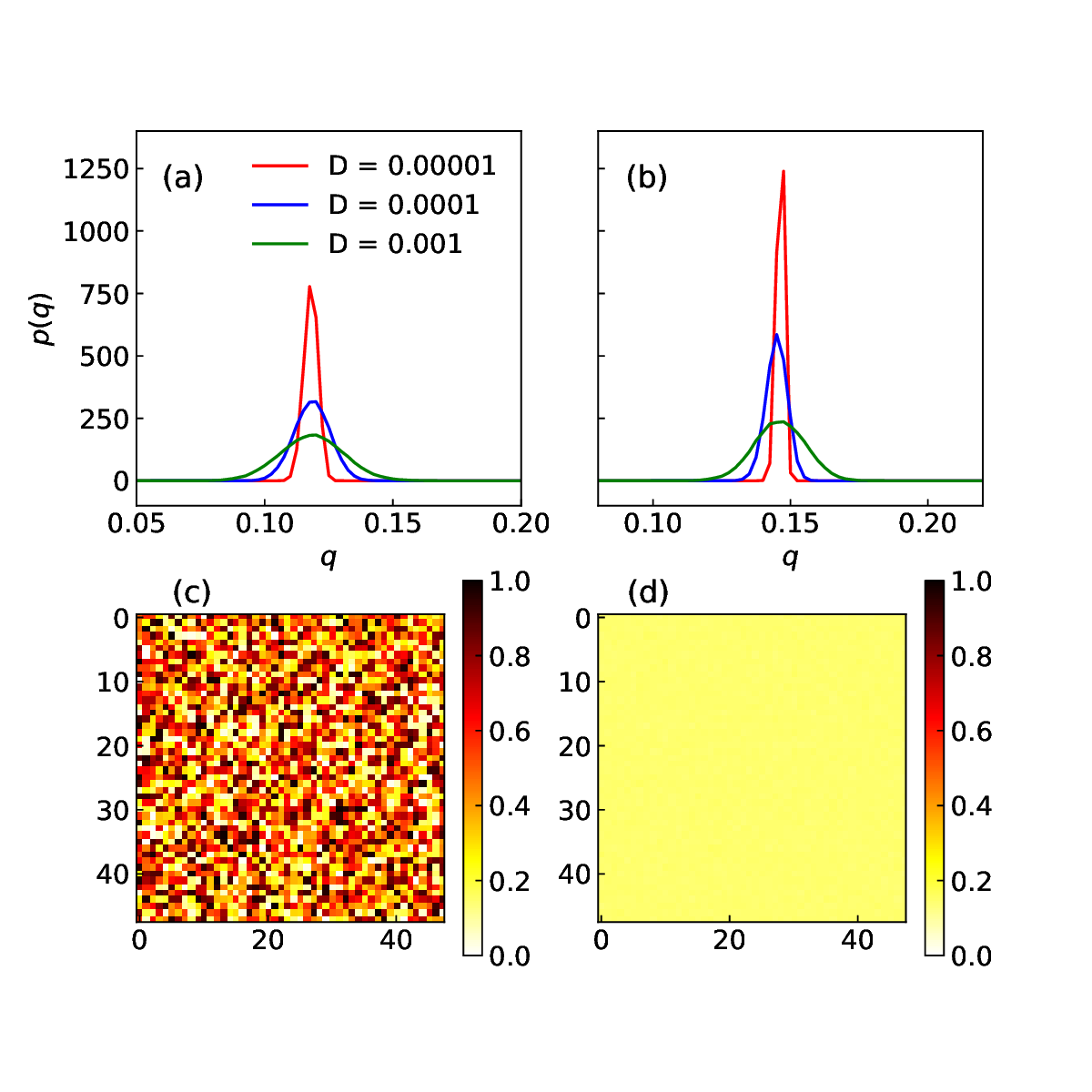}
\caption{\label{fig:distribution} Noise distributions for (a) $r =1.7$, (b) $r = 2.08$, and
the snapshot of the distribution of (c) initial state and (d) steady state for 
$D = 0.0001$ and $r = 2.08$. The color axis represents the values of the noise parameter $q_{i}$.}
\end{figure}

To begin with, we computed the distribution of noise $q$ in noise layer to ensure it remains in a stable
state during the evolution of vote layer.
The results for $r = 1.7$ and $2.08$ are illustrated in Fig. \ref{fig:distribution} (a) and (b). As the diffusion coefficient increases,
the distribution of $q$ becomes more uniform.
Fig. \ref{fig:distribution} (c) and (d) depict the initial state with a random distribution of noise $q$
and the steady state under our dynamical mechanism evolving, respectively.

\begin{figure}
    \centering
    \subfigure{\includegraphics[width=0.9\textwidth]{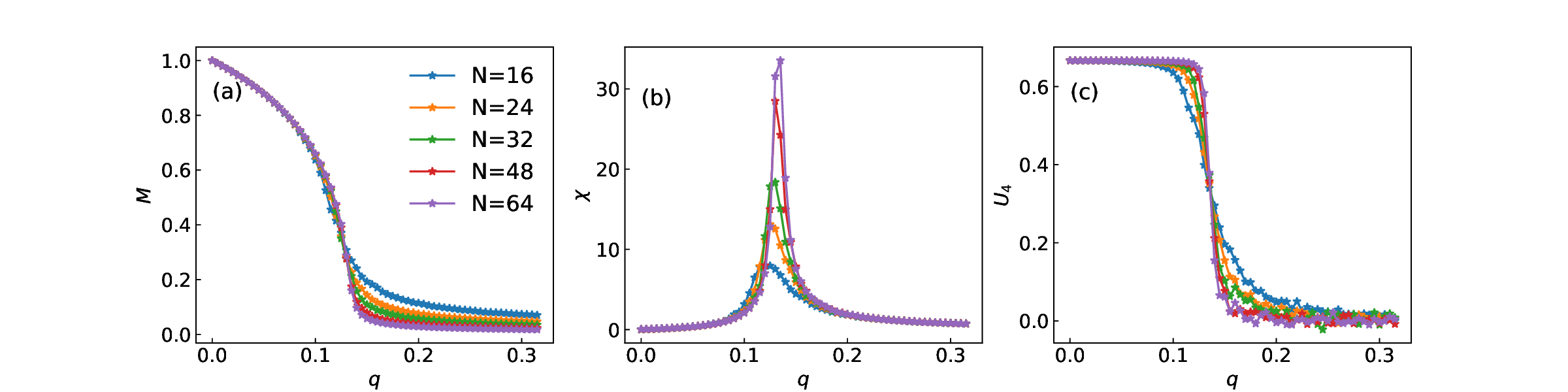}}
    \hfill
    \subfigure{\includegraphics[width=0.9\textwidth]{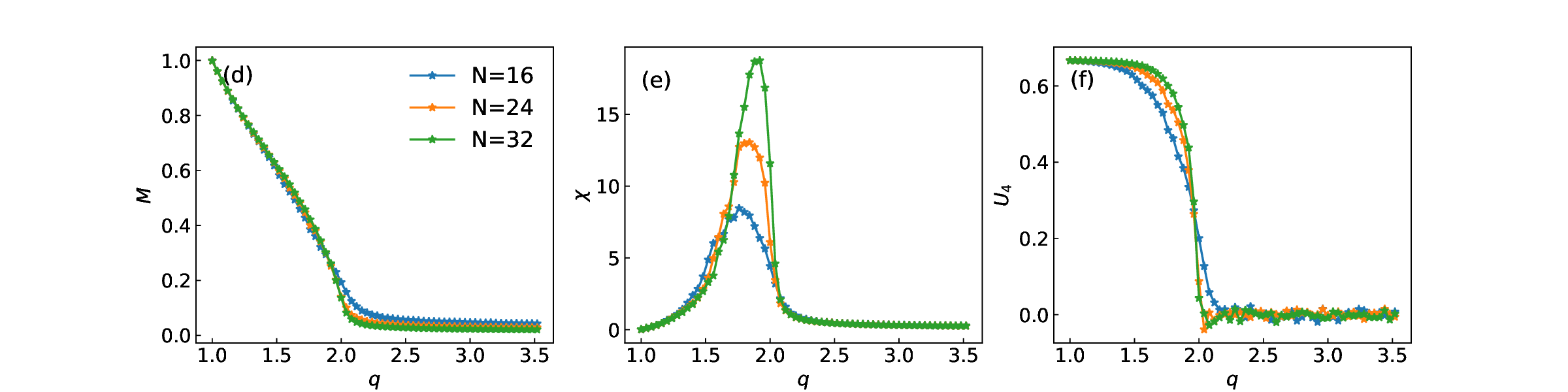}}
    \caption{\label{fig:rrnbulk}The bulk properties $M$, $\chi$ and $U_4$ as 
    a function of external noise $q$ or competing ratio$r$: MV in (a) to (c) single layer RRN with external noise, (d) to (f) multi-layer (noise layer is SL and the vote layer is RRN) with feedback.}
\end{figure}

The typical data for bulk quantities in RRNs with external noise, as well as for coupled RRN for the vote layer and square lattice for noise layer with diffusion, 
are shown in Fig. \ref{fig:rrnbulk}. 
The curves of the Binder cumulants behave as normal second-phase transitions
crossing at the fixed point.

\begin{figure}[htbp]
\centering
\includegraphics[width=0.85\textwidth]{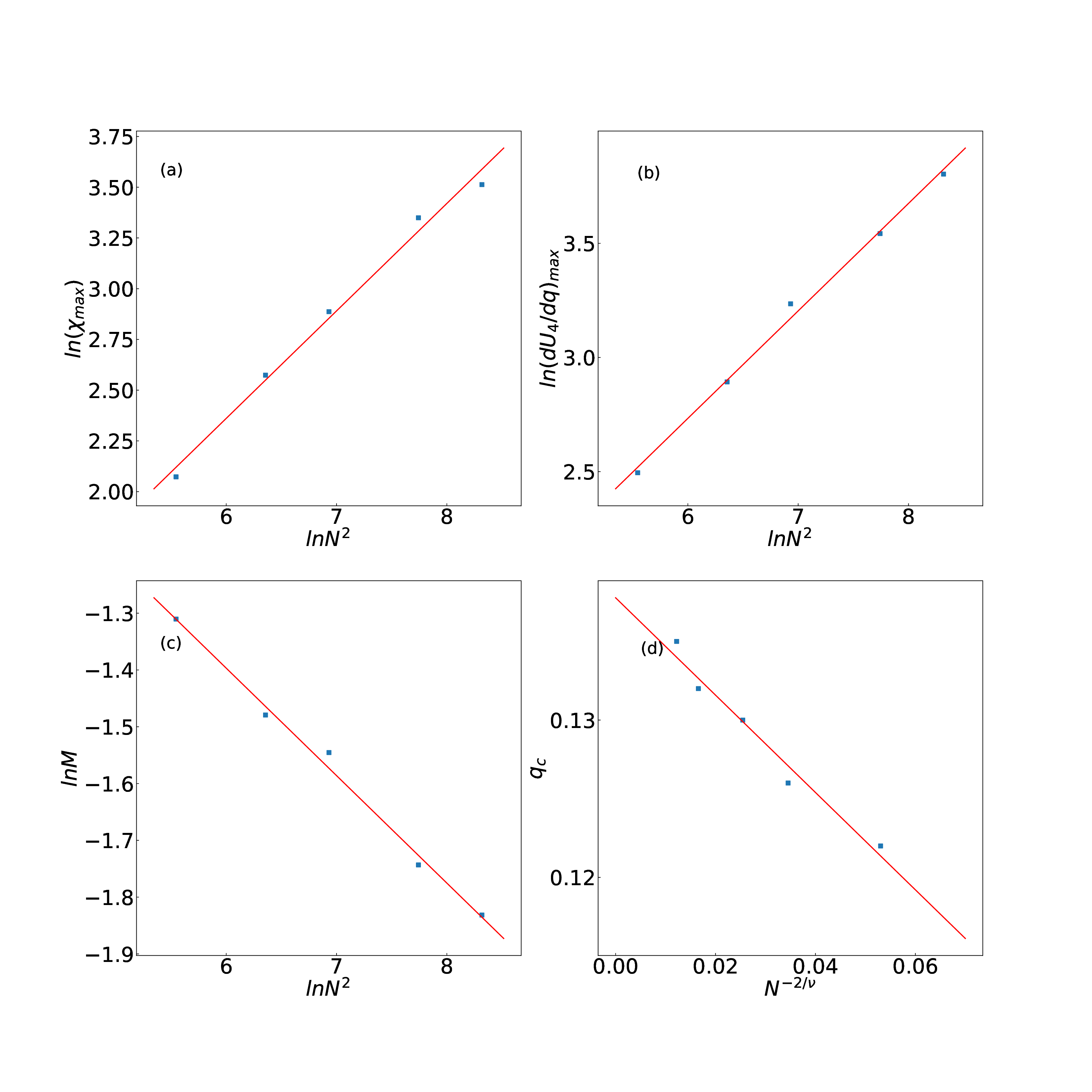}
\caption{\label{fig:rrnscaling} Finite size analysis for single RRN with external noise. Size dependence of the maxima of (a) $\chi$, (b) $\frac{dU_4}{dq}$, respectively, and (c) the value of $\ln M$ on the position of the effective transition point. (d) Effective transition noise $q_c (N)$ according to the relation scaling (\ref{criticalt}). The slope of the red lines in (a) to (c) give the critical exponents $\gamma /\nu$, $1/\nu$ and $\beta / \nu$ respectively, and the line of the intercept of the vertical axis in (d) gives the position of the critical noise.}
\end{figure}

Applying finite-size analysis, we can obtain the critical exponents and critical noise $q_c$ or critical
ratio $r_c$. Fig. \ref{fig:rrnscaling}
gives the maximum of the corresponding quantities as a function of the lattice sizes for RRN.
We plot the data of the logarithmic values and find excellent scaling behavior. 

We then calculate the other cases: 1)the single-layer networks with homogeneous
external noise, and 2)the single-layer networks with the noise feedback
from the vote layer (the noise is generated via Eq. (\ref{noise_wod})), and 3) the coupled RRNs for the vote layer(VL) and the noise layer(NL), and 4) the coupled RRN for the
VL and the SL for the NL, and 5) the coupled RRN for the
NL and the SL for the VL, and 6) the coupled SL for both layers (The noises are
generated via Eq. \ref{noise_ev} for the cases 3) to 6)). The results are consistent with the previous ones \cite{de1992isotropic,liu2019coevolution}
for the case 1) to 3). The results are in Table \ref{tab:critcal}. 

\begin{table}[ht]
\centering
\caption{\label{tab:critcal}The critical exponents and critical temperature $T_c$, ratio $r_c$
or noise $q_c$ for different networks. $z$ is the coordination
number of each site, $J$ the
exchange interaction between adjacent spins and $k_B$ Boltzmann constant.}
\begin{tabular}{ccccc}
\br
    &$\beta / \nu$&$\gamma / \nu$&$1/\nu$ & $T_c$\\ \hline
2-d Ising model \cite{pathria2011statistical} & $0.125$&$1.75$ & 1 & $2.269$ \\
Mean-field Ising model  \cite{pathria2011statistical}& $0.25$& $0.5$ & $2$ & $zJ/k_B$\\
\br
    &$\beta / \nu$&$\gamma / \nu$&$1/\nu$ & $q_c$\\ \hline
  SL without feedback&$0.125 \pm 0.007$&$1.73\pm 0.02$ &$1.01\pm 0.05$&$0.075\pm 0.002$ \\
 RRN without feedback & $0.189\pm 0.009$&$0.529\pm 0.008$ &$0.482\pm 0.023$&$0.138\pm 0.001$ \\
 \br
    &$\beta / \nu$&$\gamma / \nu$&$1/\nu$ &  $r_c$\\ \hline
 RRN with feedback only& $0.18\pm 0.02$&$0.61\pm 0.02$ &$0.41\pm 0.04$&$1.62\pm 0.03$ \\
RRNs for both layers& $0.23\pm 0.03$&$0.55\pm 0.02$ &$0.47\pm 0.03$&$2.08\pm 0.03$ \\
RRN for VL and SL for NL&$0.23 \pm 0.04$&$0.57\pm 0.03$ &$0.48\pm 0.03$&$2.03\pm 0.05$ \\
\br
\end{tabular}

\end{table}

\begin{figure}[htbp]
\centering
\includegraphics[width=1.0\textwidth]{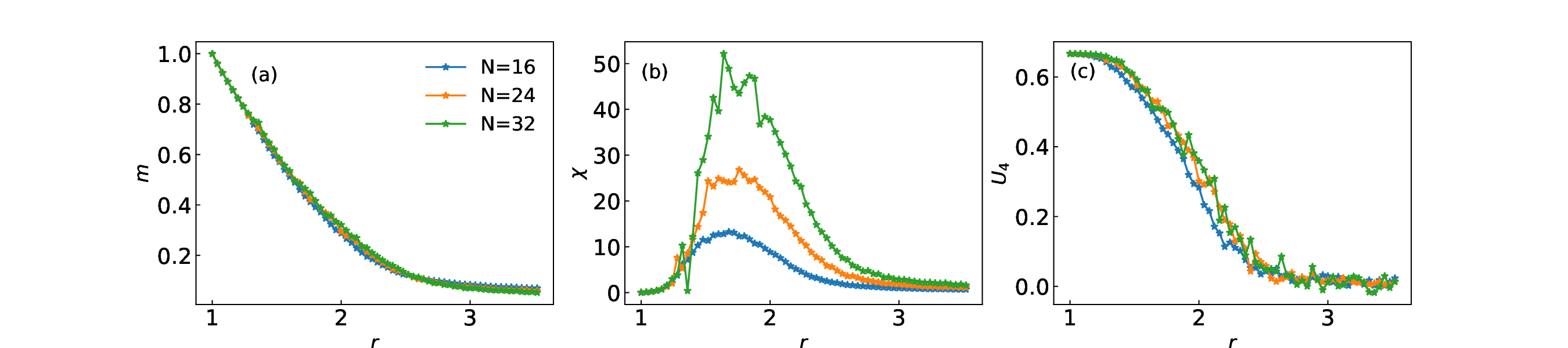}
\caption{\label{fig:squarebulk}The bulk properties $M$, $\chi$ and $U_4$ as 
    a function of competing ratio $r$ for coupled square lattice.}
\end{figure}

The critical
transitions are found in the first three cases except the SL vote layer
with noise feedback from the vote layer. We do not observe the signals of the critical transitions for the case 5) and 6). Fig. \ref{fig:squarebulk}
presents the results for the coupled square lattice. The order parameters
do not show a distinct finite-size effect and $U_4$ could not cross at a fixed
point, in addition, the width of $\chi$ does not narrow with increasing system size; instead, it widens. This suggests that as the system tends towards infinity, the response function may not necessarily exhibit divergent behavior. 

In voter model, each individual (or "voter") randomly selects a neighbor and
adopts the opinion of that neighbor. This model is commonly used to study
the process of consensus formation or the stable state of opinion
polarization.Voter behavior is commonly observed during order-disorder
non-equilibrium transitions, instigated by interfacial noise, amid
dynamically symmetric absorbing states \cite{dornic2001critical}. This symmetry can be enforced
either by an up-down symmetry within local rules or by the global
conservation of magnetization. The universal exponents associated with the
transition are $\beta = 0$ and $\nu = 1/2$ across all dimensions.
The majority-vote model is slightly more complex. In this model, individuals tend to adopt the majority opinion of their neighbors, but with a certain probability (usually represented by a noise parameter 
$q$), they make the opposite choice. Therefore, this model not only involves dynamics of local consistency but also introduces randomness, making it closer to the real-world process of opinion formation.

The model in SL without feedback is the traditional model of which the
critical exponents are the same as the 2D Ising model \cite{de1992isotropic}.
Furthermore, our observations indicate that critical transitions fail to
manifest in the square-lattice vote layer when noise originates from
feedback. Conversely, when the vote layer is a random regular network (RRN),
critical transitions consistently emerge. We examine the scaling relation
(\ref{realtion}) in the case of the RRN vote layer and discover that
$D_{eff} \sim 1$, irrespective of the presence of feedback and diffusion in
the noise layer. $1/\nu $ and $D_{eff}$ suggest that all scenarios with RRN vote layers fall into the mean-field universality classes ($\beta /\nu =0.25$, $\gamma /\nu =0.5$ and $1/\nu = 1/2$), which is
consistent with the previous work \cite{liu2019coevolution}. The research by
Sampaio Filho et al \cite{sampaio2016majority} on the majority-vote model
on spatially embedded networks shows that when the added edges tend to be
more randomly connected, the critical behavior falls into the mean field
universality class. However, when the addition of edges tends to connect
more to nearby nodes, the critical behavior is closer to that of the
2-dimensional Ising model universality class. This is similar to the results
we obtained on RRN.

\subsection{The signal of third-order transitions}

\begin{figure}[htbp]
\subfigure{\includegraphics[width=1.0\textwidth]{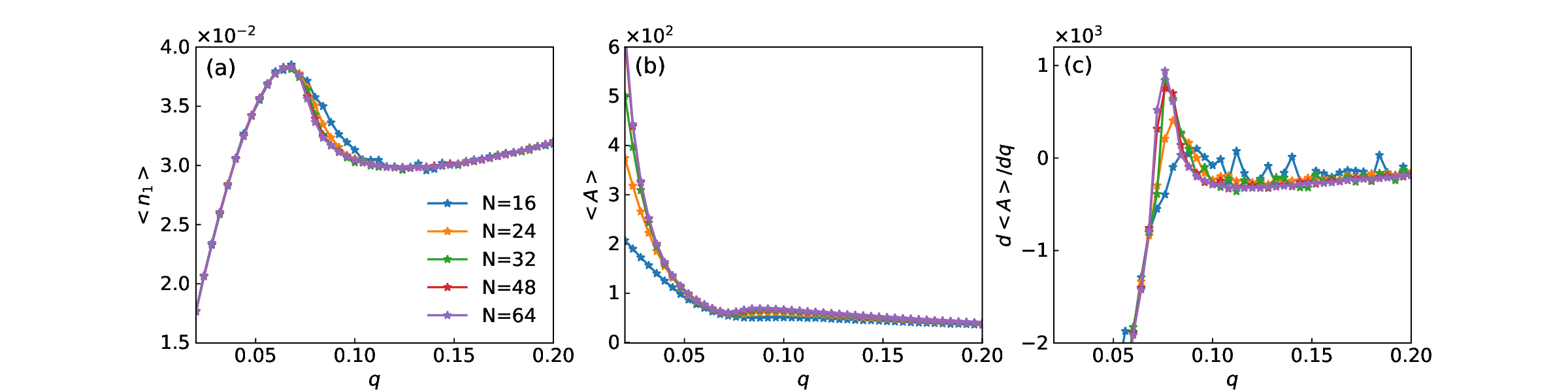}}
\hfill
\subfigure{\includegraphics[width=1.0\textwidth]{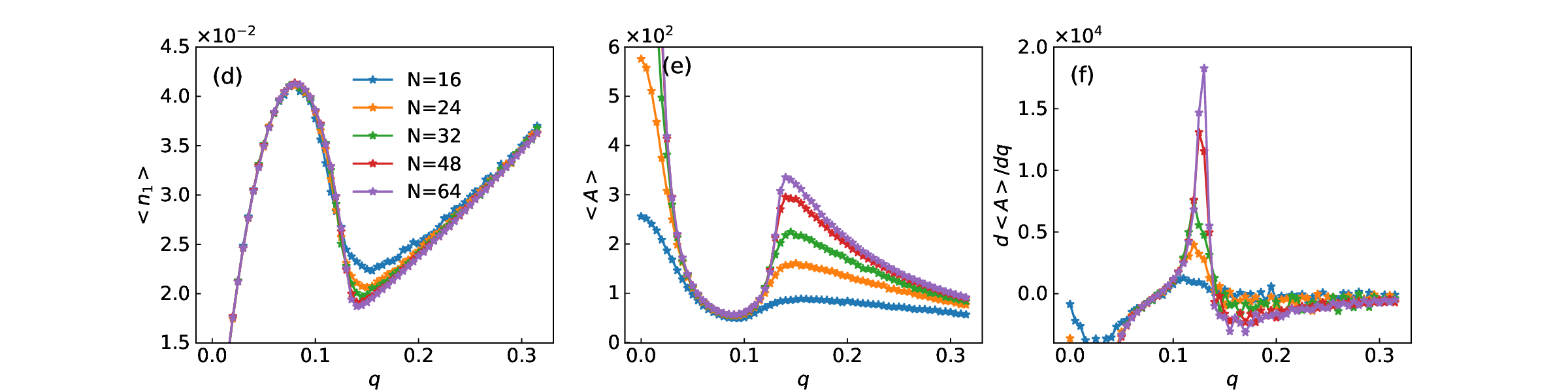}}
\caption{\label{fig:rrN3order} ISN, ACS and the derivatives of ACS for single layer square lattice (a) to (c) and RRN (d) to (f). The figure show the model with external noise $q$.
}
\end{figure}

\begin{figure}[htbp]
\subfigure{\includegraphics[width=1.0\textwidth]{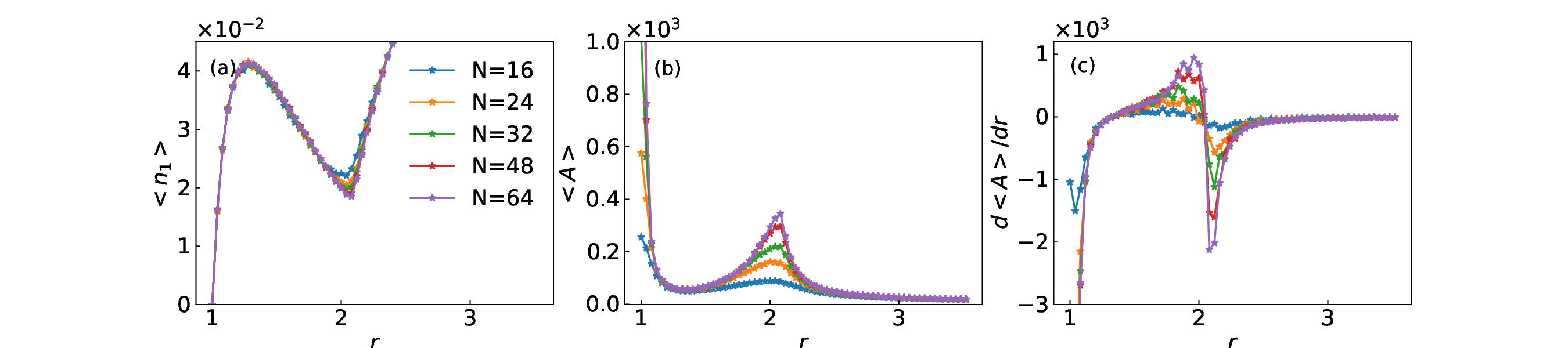}}
\hfill
\subfigure{\includegraphics[width=1.0\textwidth]{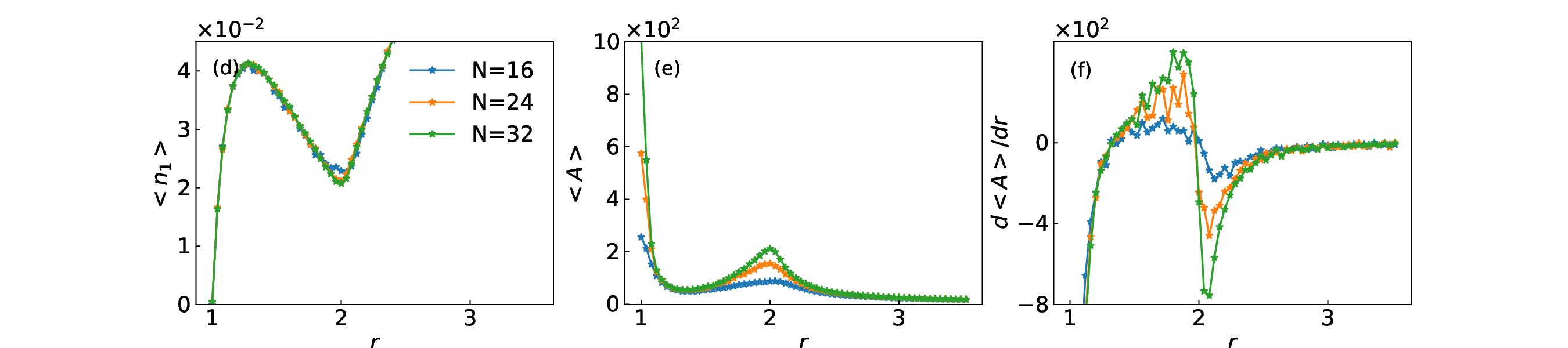}}
\caption{\label{fig:nsvrcouple3order} ISN, ACS and the derivatives of ACS as 
function of $r$. The figure shows the coupled network: (a) to (c) the vote layer and the noise layer are both RRN, while (d) to (f) the vote layer is RRNs and the noise layer is square lattices.
}
\end{figure}

\begin{figure}[htbp]
\subfigure{\includegraphics[width=1.0\textwidth]{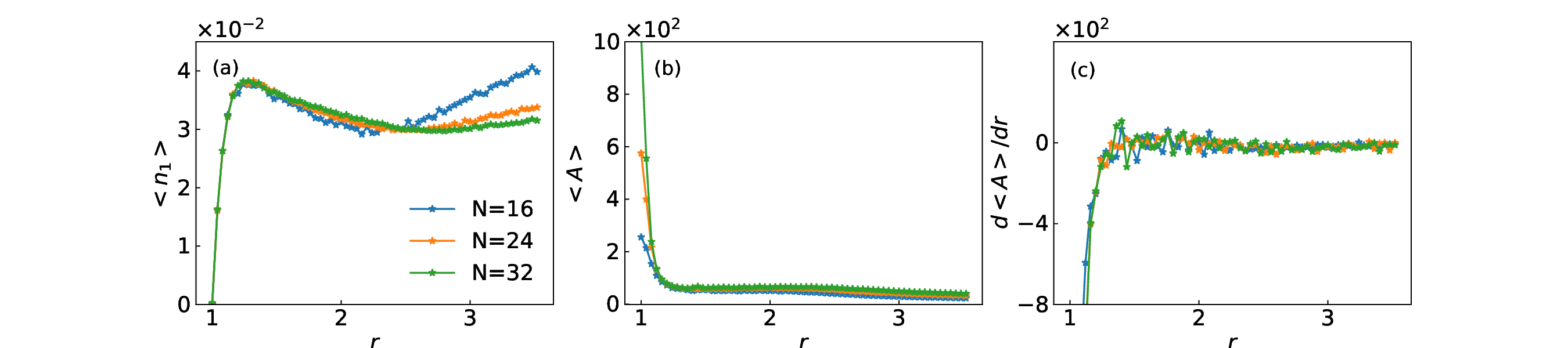}}
\hfill
\subfigure{\includegraphics[width=1.0\textwidth]{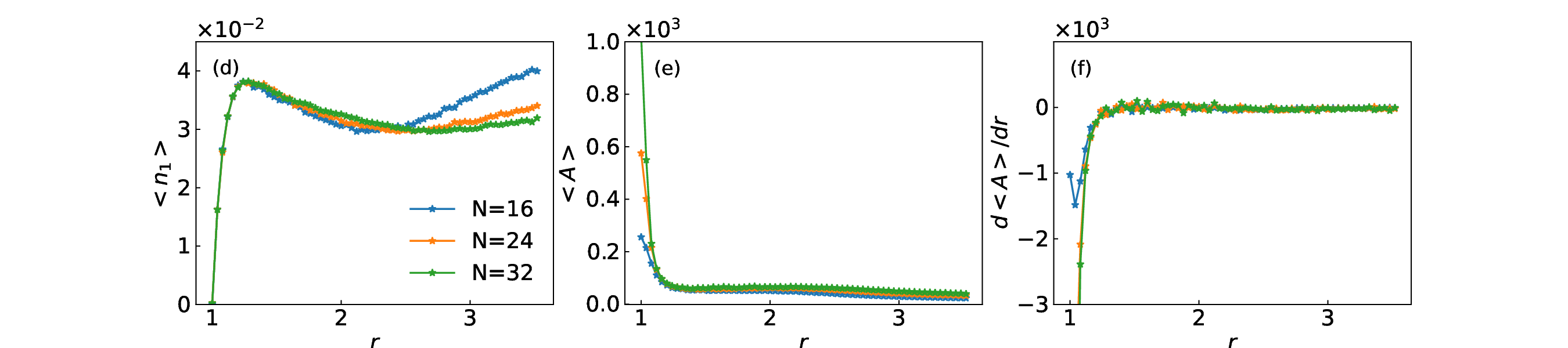}}
\caption{\label{fig:squrcouple3order}ISN, ACS and the derivatives of ACS as 
function of $r$. The figure shows the coupled network: (a) to (c) the vote layer is SL and the noise layer is RRN, while (d) to (f) the both layer are
SL. There are no distinct local minima in (d) and (f).
}
\end{figure}

The new order parameter $\langle n_1\rangle$, $\langle A\rangle$ and the derivatives of $\langle A\rangle$ are drawn
in the figures \ref{fig:rrN3order} to \ref{fig:squrcouple3order} for different
coupling networks. The curves of $\langle n_1 \rangle$ exhibit local maxima implying
independent third-order transitions.
As $q$ or $r$ increases, some systems show
critical phenomena. We can observe the region where $\langle n_1\rangle$ rapidly decreases and $\langle A \rangle$ rapidly
increases in the curves of the figures \ref{fig:rrN3order} and \ref{fig:nsvrcouple3order}.
However, if the critical transitions do not occur in some systems, the
decrease of the curves $\langle n_1 \rangle$ can be observed but they are not sharp, and the increases of
$\langle A \rangle$ cannot be observed in Fig. \ref{fig:squrcouple3order}.
While independent third-order transitions can be observed, even the systems
do not show the critical transitions in Fig. \ref{fig:squrcouple3order}.
We can see from the pictures that the positions of the third-order transitions do not depend on the size of the systems. Noise or the coupling
is the decisive factor.
Above the positions, the disorder of the systems emerges (but the systems do
not get into the disorder phase through critical phase transitions). 

The curves of the derivatives of $\langle A \rangle$ can provide signals of the dependent third-order
transition which can only exist if the first- or second-order phase transition exists \cite{qi2018classification,sitarachu2022evidence}.
Local minima are observed in figures \ref{fig:rrN3order} (c), (f) and \ref{fig:nsvrcouple3order} (c), (f).
The phenomenon is not evident in the single-layer square lattice. The reason for this
lies in the fact that local interactions lead to the slow formation of the clusters, and the
simulation noise has a noticable impact on the statistics of cluster size especially when the
system size is small. The curves of $d\langle A \rangle/dq$ when $N=48$ and $64$ in Fig. \ref{fig:rrN3order} (c) and (f) show the distinct minima, a behavior also observed in Ising model in
square lattices. For the coupled RRNs for both layers and the coupled RRN for the
vote layer and the square lattice for the noise layer, the minima of the $d\langle A \rangle/dr$ are distinct. The positions locate near and above the critical transition points.

\begin{figure}[htbp]
\centering
\includegraphics[width=0.55\textwidth]{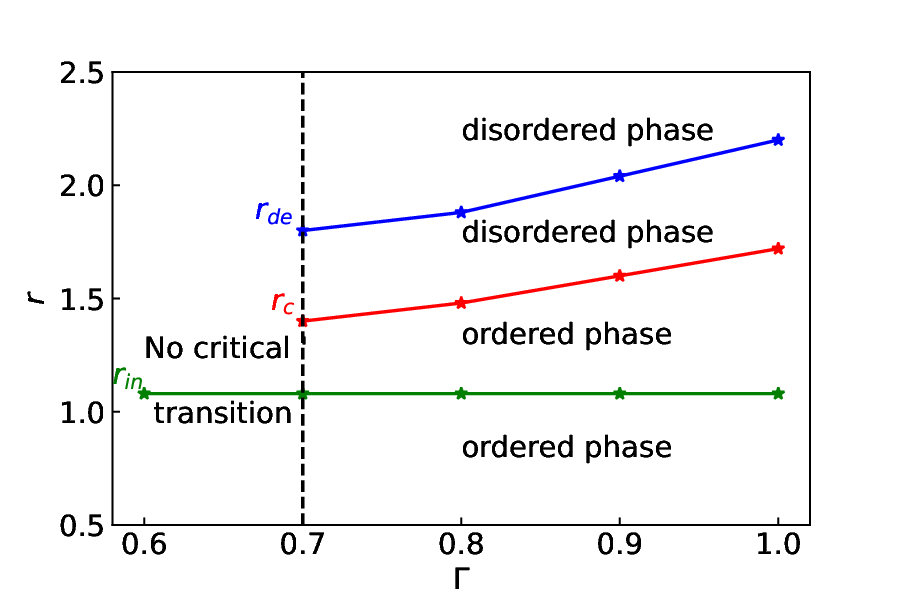}
\caption{\label{fig:phased} The phase diagram for the bilayer networks in which one layer is RRN and the other is localized RRN with different
$\Gamma$. The dashed line gives the boundary where the critical transitions disappear, the green line represents the independent third-order transition
line, the blue is the dependent third-order transition line and the red line
is the critical line.
}
\end{figure}

We then calculate the paratemetes $\langle n_1 \rangle$, $\langle A \rangle$ and $d\langle A \rangle/dr$ on the
RRN vote layers at $\Gamma = 1, 0.9,0.8,0.7, 0.6 $.
For the structure of the noise layer does not affect the critical
behaviors of the model, we use the standard RRN in this layer to accelerate the simulations.
The results are presented in Fig. \ref{fig:phased}. With the decrease of $\Gamma$,
the networks become more decentralized, and the
transition points become lower.  When $\Gamma < 0.7$, the signal of
the dependent third-order transitions vanishes. Under such circumstances, the critical transition remains unobservable through conventional finite-size analysis.

The independent and dependent third-order transitions are located on both sides of the
critical points, respectively. They provide us with a method to formulate strategies predicting the near-critical area. If $d\langle A \rangle/dr$ has a
local minimum, the first- or second-order phase transitions will occur at $r_c$.
Meanwhile, the positions of the local maxima of $\langle n_1 \rangle$ determine the other boundary
of the near-critical area. However, if $d\langle A \rangle/dr$ does not have the local minimum,
the system does not show a first- or second-order transition. $\langle n_1 \rangle$ only provides the
information of the emerging disorder.

\section{\label{sec:summary}Summary}
We have studied the transitional behaviors of the majority-vote models, beginning with an exploration of single-layer models involving external noise and noise feedback stemming from the voting clusters. We observe the critical behavior and locate the critical points for cases involving homogeneous external noise. Upon considering the feedback mechanism, the identification of a critical transition becomes elusive within the framework of a square lattice. Furthermore, we delve into the analysis of distinct noise layer structures and the propagation of noise. Notably, when a regular square lattice or a decentralized RRN with $\Gamma<0.7$ is adopted for the voting layer, critical transitions do not occur. This observation suggests that feedback instigates the absence of critical behavior for systems with decentralized interactions.

However, amidst these dynamics, a consistent pattern emerges: the presence of independent third-order transitions, indicating the onset of disorder (below a critical point), remains consistent in all scenarios we studied. In networks that can undergo critical transitions, we find dependent third-order transitions. These dependent transitions can be seen as early signs of critical transitions in discrete variable models, at the very least. By monitoring changes in the sizes of consensus clusters, it becomes possible to provide early warnings before critical phase transitions happen. This allows for timely intervention before major collective events occur, based on reliable empirical evidence.

This approach holds promise not only for anticipating critical phase transitions in discrete variable models but also for extending its potential to early warnings of critical states in other complex systems.

\section{Acknowledgements}
We acknowledge the support of NSFC (NO. 71731002 and NO. 12304257) and
the Natural Science Basic Research Program of Shaanxi (Program No. 2022JM-039)
\vspace{10pt}

\bibliographystyle{iopart-num}
\bibliography{apssamp.bib}
\end{document}